# SEARCH FOR COMPACT STELLAR GROUPS IN THE VICINITY OF IRAS SOURCES

**N. M. Azatyan, E. H. Nikoghosyan, and K. G. Khachatryan**

*The results of a search for compact clusters in the vicinity of 19 IRAS sources based on data from the GPS UKIDSS and Spitzer GLIMPSE surveys are presented. Overall, clusters have been identified in 15 regions. Clusters are identified for the first time in 4 regions (IRAS 18151-1208, IRAS 18316-0602, 18517+0437, 19110+1045). In 5 regions (IRAS 05168+3634, 05358+3543, IRAS 18507+0121, IRAS 20188+3928, IRAS 20198+3716) the compact groups we have identified are substructures within more extended clusters. The radii of the identified groups and the surface star density are widely scattered with ranges of 0.3-2.7 pc and 4-1360 stars/pc$^2$, respectively. In 11 of the clusters, the IRAS sources are associated with a pair or even a group of YSOs. The groups identified in the near IR include representatives of a later evolutionary class II among the stellar objects associated with the IRAS sources.*

**Key words:** *stars: star formation: open clusters, PMS objects, IR band*

1. **Introduction**

A set of observational data currently exists which indicates that star formation in star clusters within the galactic disk is a multistage process. Thus, objects in different stages of evolution can be observed in a given cluster.

---

V. A. Ambartsumyan Byurakan Astrophysical Observatory, Armenia;
e-mail: nayazatyan@gmail.com

It appears that the youngest formations are compact groups of young stellar objects (YSO) with diameters of ~1 pc and ages of ~1 million years embedded in a dense gas-dust medium; the central regions of these groups generally include a young active stellar object (YSO) or a pair of stars with large and medium masses [1-3]. Studies of these star formation regions may be relevant to an entire range of questions related to the evolutionary theory of individual stellar objects of different masses, as well as of clusters as a whole.

The development of long-wavelength astronomy has greatly stimulated research in this area. A number of papers have been published on searches for and studies of these kinds of star formation regions (2MASS [4-6], Spitzer GLIMPSE [7], etc.). It should be noted, however, that searches for compact clusters in the vicinity of YSOs with assumed high masses are not always successful. For example, in Ref. 5 only 57 clusters were discovered in the neighborhoods of 217 YSOs associated with IRAS sources; this represents only 26%. Thus, the question arises of why clusters were not observed in the other 74% and whether this was caused by a lack of observational data.

This paper is also a search for and determination of the sizes and numerical composition of compact clusters in the vicinities of YSOs associated with IRAS sources based on data in the near and mid IR. We have also discovered YSOs associated with IRAS sources and determined their evolutionary classes using infrared photometry data.

**2. Objects and methods**

**2.1. Objects.** In our search for compact clusters, we have chosen IRAS sources from the list of Ref. 8 which are associated with YSOs, presumably with high masses ($M > 8\ M_\odot$). Different signs of activity were observed in all regions: $H_2$ and CO flows, emission in $NH_3$, $H_2O$, and $CH_3OH$ lines, compact HII regions, etc. From this list we selected objects which lie in the region of the GPS UKIDSS near-IR survey. Overall, there are 20 regions, one of which (IRAS 05137+3919) has been examined in detail [9] where ~80 PMS stars found in a cluster with a radius of 1'.5.

**2.2. Data.** We have used images, coordinates, and photometric data from the GPS UKIDSS survey [10]. The astrometric accuracy and resolution of the survey is ~0.1 arcsec/pix. For purity of the sample with respect to the photometric parameters we have chosen objects with $K < 18^m.05$. In addition, we have excluded those objects that have a probability greater than 90% of being the result of various kinds of noise patches or image defects, as well as objects with coordinates that match densifications of directed outflows revealed in $H_2$ lines [8].

We have also used mid IR data from the Spitzer GLIMPSE survey [11]. There the resolution of the images is 0.6 and 1.2 arsec/pix. The photometric completeness of the sample for the different ranges varies from $12^m.0$ to $14^m.0$.

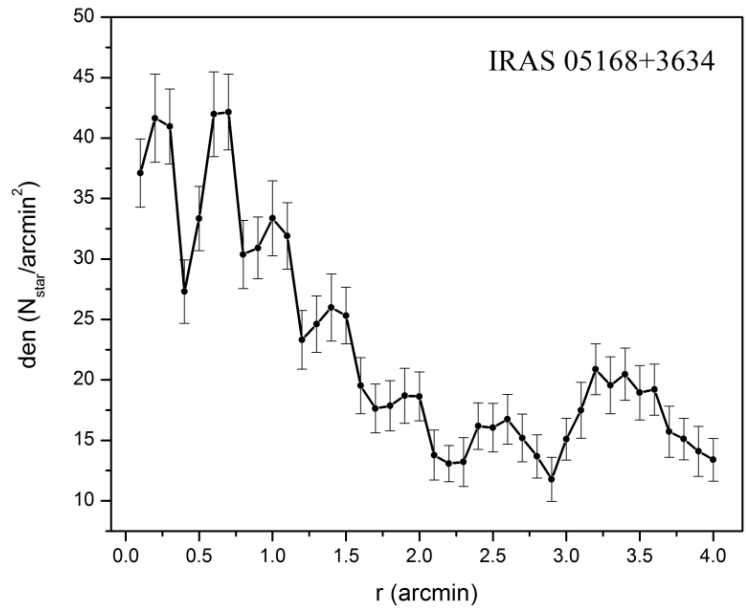

**Fig. 1.** Radial distribution of the stellar density of a cluster.

**2.3. Detection of clusters.** In order to detect clusters we have constructed maps of the distribution of surface stellar density around each IRAS source within a 4' x 4' region. The density was determined simply by dividing the number of stellar sources by the area of a square of size b x b with step size a. Three sets of values of the parameters b and a were used: 40" and 20"; 30" and 15"; and 20" and 10". In order to maximize the ability to detect a cluster, the dimensions of the square and step size were determined empirically for each cluster to increase the statistical significance of the local peaks in stellar density. A large sized square may "smear" out the local density peaks against the common background. On the other hand, if the square is too small then the overall picture is subject to error since the number of sources in each bin is comparable to the random fluctuations in the background stellar density. Ultimately, the size of the bin and step size were chosen in accordance with the size of the group itself. A group was considered to really exist if the surface density in the vicinity of the IRAS source exceeded the average background by more than 2. Isodenses of the found clusters are shown in Fig. 3.

In order to confirm the existence of clusters, and to refine their dimensions, we also constructed the radial distribution of the density relative to the geometric center of a group. The stellar density was determined for each annulus of width 0'.1 by simply dividing the number of stars by the surface area. A measure of uncertainty was obtained in accordance with Poisson statistics for the number of stars in each annulus. The radius of the cluster was taken to be the value the distance from the center beyond which the fluctuations in the stellar density in the rings become random according to Poisson statistics with a probability greater than 1%. Figure 1 is an example of the radial density distribution of a star cluster associated with the source IRAS 05168+3634. This plot shows clearly that after a distance of 1'.5 the density of the cluster does not exceed the average density of the field and that the objects in the cluster are distributed nonuniformly, forming subgroups which correspond to the shape of the isodenses in Fig. 3. Table 1 lists the radii (R) of the discovered clusters together with their "richness" (N), which was determined by subtracting the average stellar density of the field in the rings with radii from 3R to 6R from the surface stellar density in the cluster. The uncertainty in this case was determined by the relative fluctuations in the stellar density of the field.

In order to study the nature of the groups of stellar objects found inside a region, we examined their position on a *JH/HK* diagram. Conversion from the UKIRT to the CIT photometric system was done using the formulas in Ref. 16. As an example, Fig. 2 is a *JH/HK* two-color diagram plotted for objects in the cluster identified in the vicinity of the source IRAS 05168+3634. The diagram shows that a significant number of the objects lie to the right of the reddening vectors and lie in the region of Ae/Be Herbig and TTau stars, as well as YSOs of evolutionary class I, i.e., they have a significant infrared excess which cannot be explained solely by interstellar absorption and, therefore, are probable candidate PMS stars. This is further evidence of the presence of a young stellar cluster in the vicinity of the IRAS source. The number of probable candidate PMS objects for all the regions ($N_{PMS}$, see Table 1) have been estimated in similar fashion. It should be noted that in some cases it even exceeds the excess number (N) of stars relative to the average stellar density of the field. For those regions in which groups of young stars were identified in the mid IR, the colors obtained from Refs. 17 and 18 were used to select the young stellar sources.

**TABLE 1.** Major Parameters of the Compact Groups

| N (1) | IRAS (2) | Dist (kps) (3) | Av (4) | R' (pc) (5) | $N_{obj}$ (6) | $N_{PMS}$ (7) | Den (star/pc$^2$) (8) | $P_{KS}$ (9) |
|---|---|---|---|---|---|---|---|---|
| GPS UKIDSS | | | | | | | | |
| 1 | 05168+3634 | 6.1 | 3.6 | 1.5 (2.7) | 93 | 80 | 4.1 | 0.003 |
| 2 | 05358+3543 | 1.8 | 5.4 | 1.2 (0.6) | 154 | 73 | 136 | 0.007 |
| 3 | 18151−1208 | 3 | 2.8 | 0.3 (0.3) | 20 | 14 | 71 | 0.027 |
| 4 | 18316-0602 | 3.2 | 3.9 | 0.5 (0.5) | 53 | 38 | 68 | 0.001 |
| 5 | 18507+0121 | 3.9 | 5.3 | 0.4 (0.5) | 45 | 13 | 57 | 0.010 |
| 6 | 19110+1045 | 6, 8.3 | 2.9 | 0.3 (0.5, 0.7) | 52 | 53 | 66, 34 | 0.469 |
| 7 | 19213+1723 | 4.3 | 0.5 | 0.3 (0.4) | 19 | 30 | 38 | 0.000 |
| 8 | 19388+2357 | 4.3 | 3.0 | 0.3 (0.4) | 7 | 24 | 14 | 0.019 |
| 9 | 20056+3350 | 1.7 | 0.3 | 0.6 (0.3) | 93 | 53 | 329 | 0.041 |
| 10 | 20188+3928 | 3.9 | 5.1 | 0.5 (0.6) | 13 | - | 12 | 0.031 |
| 11 | 20198+3716 | 0.9, 5.5 | 2.6 | 1.3 (0.3, 2.1) | 385 | 298 | 1362, 28 | 0.001 |
| GLIMPSE | | | | | | | | |
| 12 | 18360-0537 | 6.3 | 2.8 | 0.4 (0.7) | 13 | 12 | 8.4 | 0.039 |
| 13 | 18517+0437 | 2.9 | 5.6 | 0.3 (0.3) | 4 | 5 | 14 | 0.022 |
| 14 | 19374+2352 | 4.3 | 2.6 | 0.8 (1.0) | 25 | 8 | 8.0 | 0.039 |
| 15 | 19092+0841 | 4.5 | 2.6 | 0.4 (0.5) | 7 | 5 | 8.9 | 0.136 |
| No cluster | | | | | | | | |
| 16 | 18174-1612 | 2.1 | 7.3 | | | | | |
| 17 | 18385-0512 | 2.1 | 2.2 | | | | | |
| 18 | 19410+2336 | 2.1, 6.4 | 4.4 | | | | | |
| 19 | 20126+4104 | 1.7 | 4.9 | | | | | |

(1) sequence number, (2) IRAS source, (3) distance (in those cases where the distances to the source have been estimated more than once, both values are given), (4) absorption by the interstellar medium in the direction toward the region [19,20], (5) radius of group, (6) excess number of objects, (7) number of probable PMS stars, (8) surface density defined relative to the excess number of objects, (9) measure of consistency between the luminosity functions in the group and in the field.

We have also compared the luminosity functions (LF) of the stellar objects lying in the region of a cluster and in the field. For this purpose, we have compared two samples: (1) objects lying in the region of the cluster and objects lying in a ring with radii 3R and 6R. The number of stars in the second sample was reduced to the area of the cluster. The measure of consistency between the luminosity functions for these two samples ($P_{ks}$, Table 1) was determined using the Kolmogorov-Smirnov test. Note that the luminosity function was constructed in the same range as that in which the compact groups were discovered.

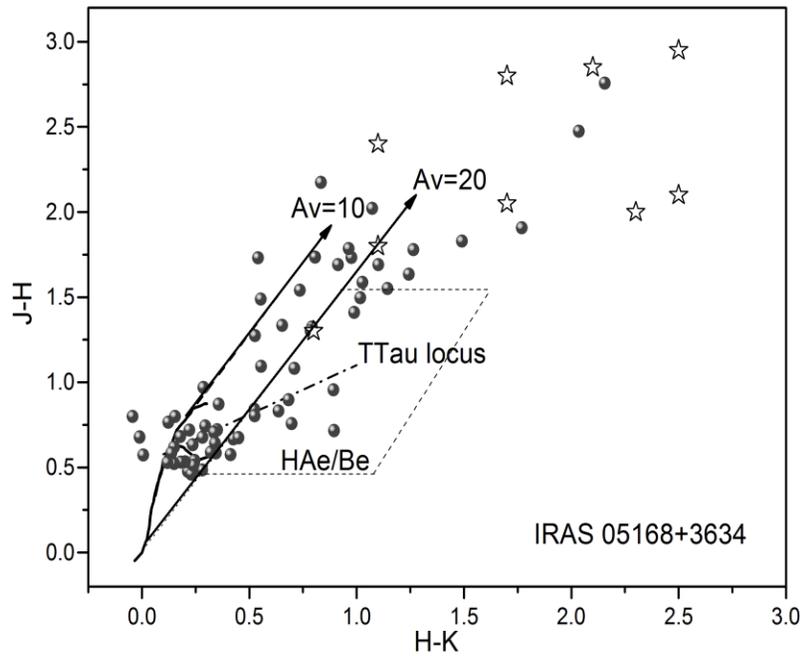

**Fig. 2.** JH/HK two-color diagram. Shown are the position of the MS and GB [12], TTau locus [13], position of the Ae/Be Herbig stars and objects of evolutionary class I (stars) [14], and reddening vectors [15].

### 3. Results
**Results of the search for compact groups.** The search method described above for compact groups in the vicinity of IRAS sources was applied to data in the near and mid IR. The results are listed in Table 1. The isodenses of these regions are shown in Fig. 3, which contains images of the regions with the positions of the IRAS and MSX sources indicated, along with the YSOs associated with them.

We were able to detect compact groups in 12 regions based on near IR data. In four cases groups were found using data from the longer wavelength mid IR range. Overall, this represents 80% of the regions that were examined. The radii and stellar density have large scatter. In one region (IRAS 20188+3928) we were unable to detect stars with distinct IR excesses. It should be noted, however, that a certain number of young stellar objects undoubtedly lie between the reddening vectors and our value of $N_{PMS}$ is only a lower bound on the overall number of PMS stars in this region. In two regions (IRAS 19110+1045 and IRAS 19092+0841), the agreement between the luminosity functions of the cluster and field is better than 10%. Nevertheless, other criteria (elevated surface stellar density, existence of a substantial number of objects with a distinct IR excess) imply quite reliably that the detected group of young stars is a real formation. In two cases (IRAS 05168+3634 and IRAS 05358+3543), the clusters that we have identified have a bimodal structure, as shown in Fig. 3. The subgroups located directly in the vicinities of the IRAS sources have radii of 0.5 and 0.2 pc and surface densities of 10 and 199 stars/pc$^2$, respectively; this exceeds the density in the cluster as a whole. We were unable to detect compact groups in four of the regions.

The data in Table 1 show clearly that there are no particular differences in the distances or interstellar absorption for the regions in which a compact group was detected in the near IR (1-11) and mid IR (12-15) or where we were unable to detect a group (16-19).

**YSOs associated with an IRAS source.** Table 2 lists the parameters of the young stellar objects associated with IRAS sources and which are the central objects in the detected compact groups.

**IRAS 05168+3634.** This is a bimodal cluster; one of the subgroups is concentrated around the IRAS sources. It is associated with an YSO in evolutionary class I. Our results differ somewhat from similar earlier work. The cluster in this region has already been identified in a 2MASS image [5]. That isodenses is similar in shape to ours, but the number of proposed members of the cluster is considerably smaller, most likely because of the difference in the photometric limit for the data. Only the region within a radius of ~20" immediately surrounding the IRAS source is examined in Ref. 6.

**TABLE 2.** Parameters of the Central YSOs

| N (1) | YSO (2) | Δ1(″) (3) | Δ2(″) (4) | Δ3(″) (5) | $\alpha_{MSX}$ (7) | $\alpha_{IRAS}$ (8) | Classification (9) |
|---|---|---|---|---|---|---|---|
| | GPS UKIDSS | | | | | | |
| 1 | J05201643+3637186[1] | 4.3 | 0.1 | 1.1 | 2.1 | 2.6 | Class I |
| 2 | J05390992+3545172[1] | 7.3 | 4.0 | 2.0 | 2.8 | 2.2 | HAe/Be |
| 3 | J18173150-1206179[1] | 15.5 | 0.5 | 0.6 | 2.9 | 2.5 | Class I |
| 4 | J18342090-0559458[1] | 16.5 | 1.6 | 1.9 | 2.4 | 2.6 | Class I |
| 5 | | | | | 3.6 | 2.8 | group |
| | J18531775+0124547[1] (B) | 5.3 | 12.9 | 13.7 | | | HAe/Be |
| | G034.4043+00.2297[2] (C) | 14.3 | 5.1 | 6.1 | | | Class I |
| | G034.4035+00.2287[2] (D) | 16.7 | 1.3 | 2.1 | | | Class I |
| | G034.4032+00.2279[2] (E) | 19.6 | 2.4 | 1.4 | | | b, Class I |
| 6 | J19132208+1050538[1] | 2.3 | 3.5 | 1.3 | 3.7 | 1.5 | b, Class I |
| 7 | J19233728+1729024[1] | 4.4 | 0.6 | 1.0 | 3.0 | 1.5 | HAe/Be |
| 8 | G059.8329+00.6719[2] | 5.3 | 31.7 | 3.3 | 3.1 | 2.0 | b, ? |
| 9 | J200731.37+335940.9[3] | 2.7 | 2.8 | 4.0 | 2.3 | 2.1 | b, Class I |
| 10 | | | | | 4.7 | 1.6 | group |
| | J20203907+3937586[1] | 7.5 | 5.4 | 6.0 | | | Class I |
| | J20203934+3937552[1] | 3.3 | 2.1 | 2.9 | | | Class I |
| | J20203902+3937531[1] | 4.3 | 3.4 | 2.9 | | | Class I |
| 11 | | | | | 3.7 | 1.5 | group |
| | J20214157+3726061[1] | 14.3 | 16.1 | 14.6 | | | Class I |
| | J20214129+3726057[1] | 12.8 | 15.6 | 6.0 | | | Class I |
| | J20214128+3725559[1] | 3.2 | 6.2 | 2.9 | | | HAe/Be ? |
| | J20214085+3725359[1] | 17.8 | 16.8 | 2.9 | | | HAe/Be |
| | GLIPMSE | | | | | | |
| 12 | | | | | 3.7 | 2.5 | pair |
| | G026.5123+00.2822[2] | 15.8 | 9.0 | 5.6 | | | Class I |
| | G026.5116+00.2844[2] | 14.3 | 11.3 | 8.0 | | | Class I |
| 13 | J18541340+0441210[1] | 12.5 | 36.2 | 3.5 | 2.2 | 2.9 | ? |
| 14 | G059.6024+00.9119[2] | 20.8 | 1.2 | 3.5 | 3.0 | 2.2 | b, Class I |
| 15 | G043.0377-00.4513[2] | 18.4 | 16.9 | 14.6 | 1.5 | 3.1 | Class I |
| | No cluster | | | | | | |
| 16 | | | | | 4.4 | 2.8 | |
| 17 | J184113.21-050900.3[3] | 19.4 | 39.6 | 1.1 | 3.7 | 2.4 | t, Class I |
| 18 | | | | | 3.3 | 2.0 | group |
| | J19431084+2344047[1](B) | 11.5 | 5.3 | 6.6 | | | Class I |
| | J19431121+2344039[1](A) | 6.1 | 0.6 | 1.3 | | | b, Class I |
| | J19431120+2344117[1](C) | 8.3 | 7.3 | 7.0 | | | Class I |
| 19 | J201426.10+411331.5[3] | 2.1 | 0.0 | 3.2 | 5.0 | 1.6 | Class I |

(1) sequence number of IRAS source (see Table 1), (2) designation of point source in the 2MASS[1], GLIMPSE[2], WISE[3] data bases, (3-5) displacement of YSO from IRAS source from corrected coordinates of IRAS source [21] and from MSX source, (6,7) the slope $(\log \lambda_1 F_{\lambda 1} - \log \lambda_2 F_{\lambda 2})/(\log \lambda_1 - \log \lambda_2)$ in the MSX and IRAS bands (60 - 100 $\mu_m$), (8) classification of the object with respect to its position on the two-color diagram (b- binary, t- triplet).

**IRAS 05358+3543.** This is the already known binary cluster Sh 2-2332RNE and Sh 2-2331RSW. The radius 1'.1 that we found corresponds to the region of the cluster where the star formation efficiency (SFE, the ratio of the stellar mass to the total mass of stars and gas) exceeds 20% [22]. At distances exceeding 1'1, SFE ≈ 5%, which is below the minimum for star formation regions [2]. In the *JH/HK* diagram the central object lies in the region of Ae/Be Herbig stars.

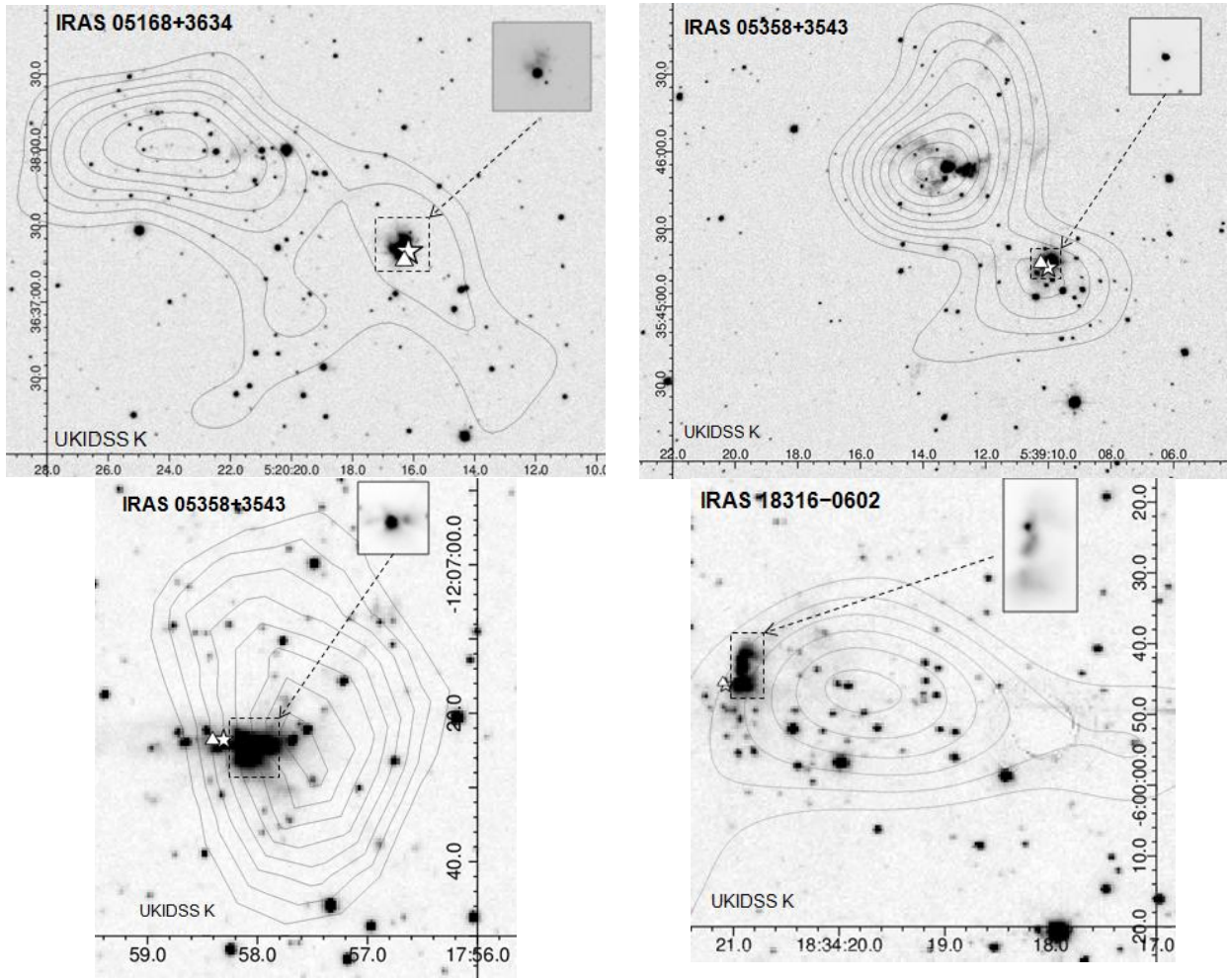

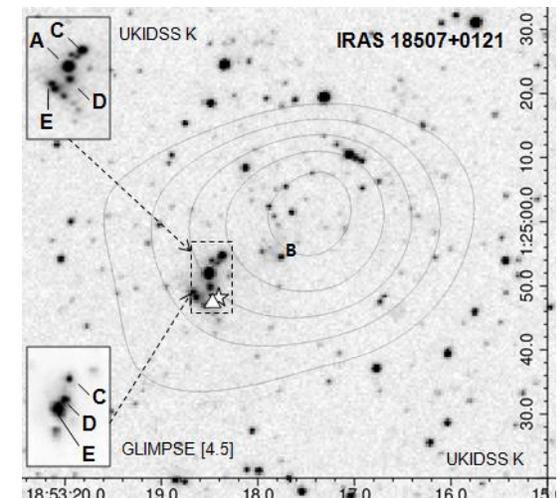
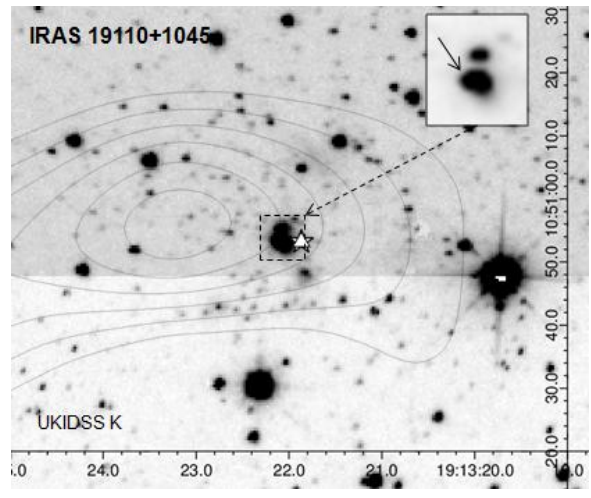
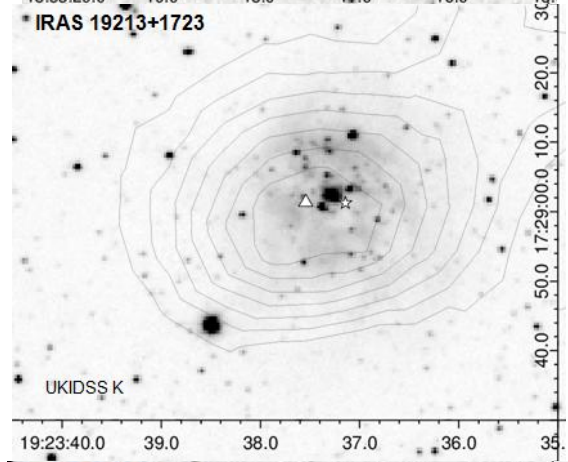
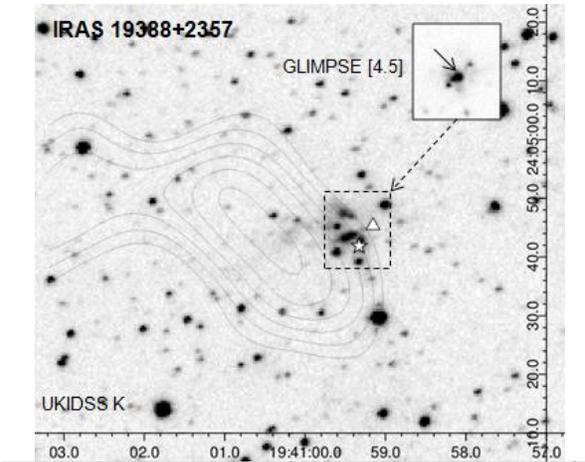
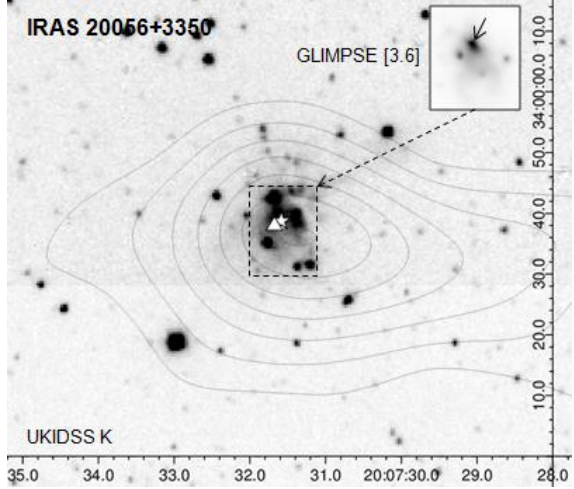
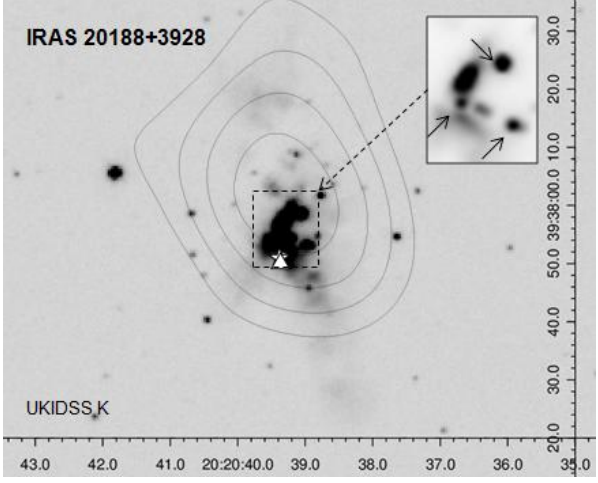

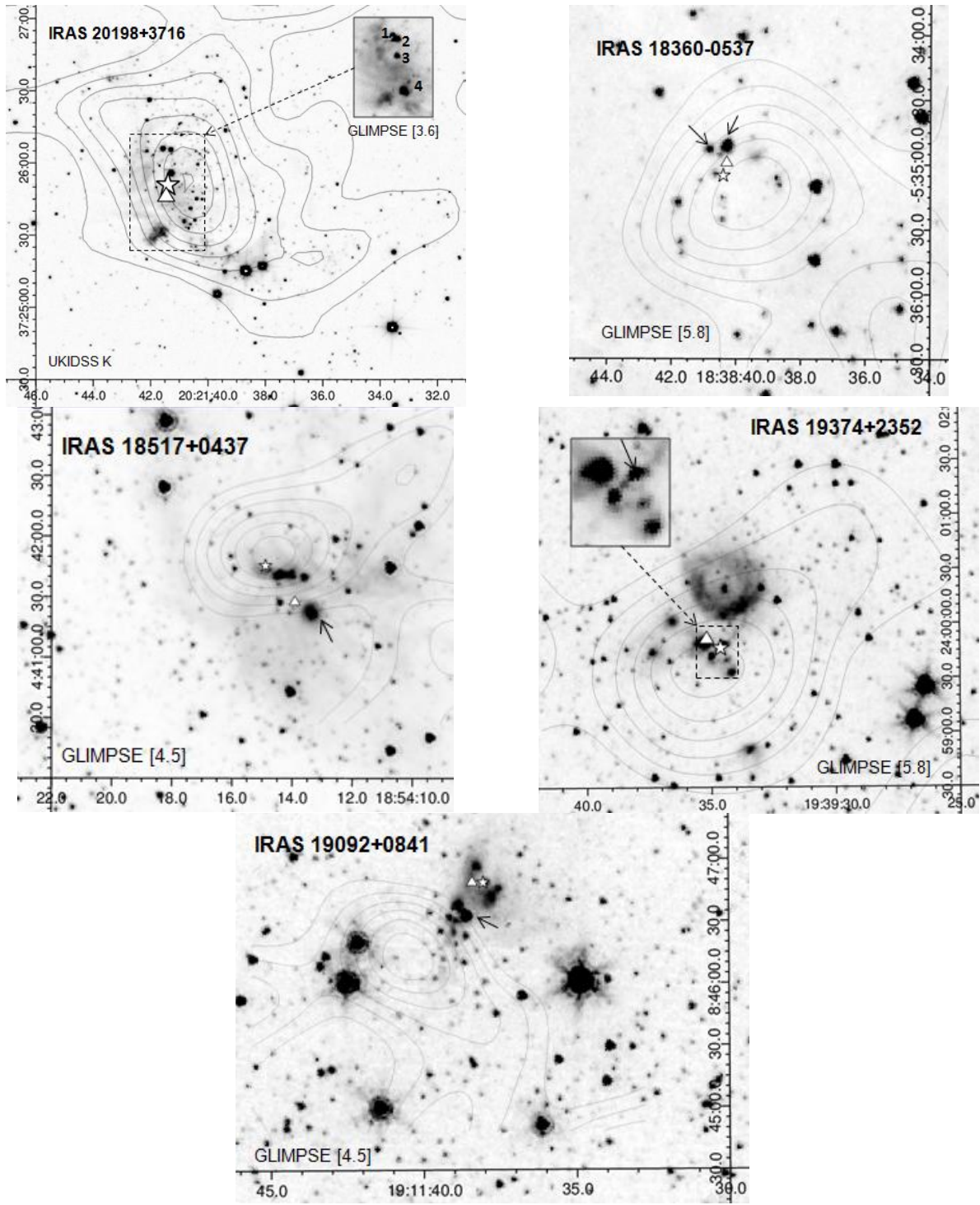

**Fig. 3.** Images and isodenses of the identified clusters. ★ - IRAS, △ -MSX sources.

**IRAS 18151-1208.** This previously unknown compact group of young stars is associated with the HI shell GSH 018+02+27 [23]. The central object has a substantial IR excess and appears to belong to evolutionary class I. In previous papers the ZAMS Sp of this young star was estimated to be B0.

**IRAS 18316-0602.** This previously unknown compact group of young stars is lays in the ionized HII region RAFGL 7009S. As in the previous case, the central object has a very substantial IR excess and can be assigned to evolutionary class I. In previous papers the ZAMS Sp of this young star was also estimated to be B0.

**IRAS 18507+0121.** Infrared stellar objects with estimated ages below $3 \cdot 10^6$ years have been discovered previously in this region [24]. They have also been identified in the mid IR (G3CC 61 [7]). The two stars "A" and "B" (see Fig. 3) have been proposed [8] as candidate YSOs associated with the IRAS. According to the 2MASS data, only object "B" has an IR excess, according to which it can be assigned to the Ae/Be Herbig objects. In addition, according to data from the GLIMPSE survey, in this region there are at least another three YSOs ("C," "D," "E") which could be classified as in evolutionary class I. Object "E" is a binary. It is entirely possible that the IRAS and MSX sources associated with objects "D" and "E," to which they are very close. Object "C" is of special interest. It is indistinguishable in the 2MASS image and image K of Ref. 8, but has substantial brightness in the GPS UKIDSS and GLIMPSE surveys. This object may be a previously unknown eruptive variable.

**IRAS 19110+1045**. This is a previously unknown compact group of young stars. The IRAS source is associated with a YSO that has a substantial IR excess. The isophotes of the star in the UKIDSS K images are somewhat elongated, which is indicative of a binary.

**IRAS 19213+1723**. This compact cluster has already been identified in the near IR [6]. However, our value for the radius is somewhat smaller. The IRAS and MSX sources appear to be associated with an object of evolutionary class II.

**IRAS 19388+2357**. As in the previous case, this group has also already been identified in the near IR [6]. The most likely YSO associated with the IRAS source shows up only in the longer wavelength mid IR (see Fig. 3). This is a close pair of YSOs for which the rising brightness in the images in the longer wavelength range greatly exceeds that of the other objects in the cluster. Photometric data were taken only in ranges (5,8) and (8,0), which is insufficient for classifying them.

**IRAS 20056+3350**. This cluster has already been included in a list of assumed stellar clusters discovered by Gaussian filtration of the UKIDSS survey (G071.312+00.827 [25]). We have obtained a refined value for its radius and determined the number of components. The YSO associated with the IRAS in the near IR is distinguishable only in the K band, where its binary character is also quite evident. According to the WISE photometric data, it can be assigned to the very young stellar objects of evolutionary class I [26].

**IRAS 20188+3928**. This group is the central part of the cluster G077.46+01.76 [4], which was discovered in images from the 2MASS survey. At least three YSOs of evolutionary class I lie in the vicinity of the IRAS source.

**IRAS 20198+3716**. This, the richest in our list of groups of young stars, is the central core of the cluster Berkeley 87, which has an age of ~$1.2 \cdot 10^6$ years [27]. Of the YSOs in the vicinity of the IRAS which stood out for their brightness in the Spitzer images we can distinguish 4 objects embedded in the gas-dust matter (see Fig. 3). Of these, No. 3, which lies closest to the source embedded in the IRAS according to the 2MASS survey data, has the smallest IR excess. Objects "1" and "2" can be assigned to evolutionary class I. It is entirely possible that the group of these young stars as a whole is the IRAS source.

**IRAS 18360-0537**. This group was also identified in Ref. 7 (G3CC 59) as a deeply embedded cluster. Two YSOs, which can be assigned to class I according to the photometry data, are closest to the IRAS source

**IRAS 18517+0437**. This, the most rarefied in our list of groups of PMS objects, is identified here for the first time. The brightest object, surrounded by a spherical nebula, is closest to the IRAS and does not have an IR excess according to the near IR data. It can be assigned to evolutionary class III. It should be noted that, on the other hand, the refined coordinates of the IRAS [21] in this region show it is further from the above mentioned star. It may be assumed that its source is a YSO deeply embedded in gas-dust matter which is indistinguishable in the IR images.

**IRAS 19374+2352**. A group of young stars has previously been discovered in this region [6], but we were not able to identify it in the near IR. This group is, however, well resolved in the mid IR (see Fig. 3). A star in evolutionary class I is associated with the IRAS source. Its elongated isophote suggests that it is a binary object.

**IRAS 19092+0841**. The group of young stellar objects in this region was also identified in Ref. 6 and classified as a deeply embedded cluster. The object associated with the IRAS and MSX sources is still ambiguous. In the GLIMPSE images, the brightest point object with an IR excess is a star in evolutionary class I (see Fig. 3) according to the photometric data. Immediately adjacent to the IRAS and MSX sources, however, there is a bright infrared nebula, within the vicinity of which we were unable to identify any point objects. It is entirely possible that a deeply embedded YSO that is not distinguishable in the images lies in this region.

**IRAS 18174-1612**. We were unable to identify any stellar objects associated with the IRAS source in this region.

**IRAS 18385-0512**. Immediately adjacent to the MSX, but at a considerable distance from the IRAS, there is a triplet of young stellar objects which can be assigned to evolutionary class I based on the WISE data.

**IRAS 19410+2336**. We were unable to discover a region with elevated surface density in the neighborhood of this source. However, a deeply embedded stellar cluster was identified in Ref. 7. Three YSOs of evolutionary class I are immediately adjacent to the MSX and IRAS sources. Here object "A" is binary. The notation for the YSOs is the same as in Ref. 8.

**IRAS 20126+4104**. A young stellar object of evolutionary class I is associated with the MSX and IRAS.

## 4. Discussion and conclusion

A statistical analysis of the star population in the vicinity of 19 IRAS sources has revealed compact groups of PMS objects in 11 regions according to near IR data and in 4 regions according to mid IR data. This represents 80% of the overall number of regions that were studied and is substantially greater than the results based on data from the 2MASS survey (see the Introduction). In four regions (IRAS 18151-1208, 18316-0602, 18517+0437, and 19110+1045) clusters were discovered for the first time. In five regions (IRAS 05168+3634, 05358+3543, 18507+0121, 20188+3928, and 20198+3716) the compact groups which we have identified are substructures that form part of more extended clusters and presumably are regions of secondary (or higher) star formation waves. We were unable to discover a compact group in the vicinities of 4 IRAS sources (IRAS 18174-1612, 18385-0512, 19410+2336, 20126+4104). However, the sources IRAS 18385-0512 and 19410+2336 are associated, respectively, with a triplet of YSO and with at least four YSOs, which can essentially be treated as a very sparse group. In the case of IRAS 20126+4104, the YSO has the highest slope a for an MSX source, i.e., it is presumably in a very early stage of evolution and is surrounded by a massive gas-dust shell which causes considerable absorption in the immediate vicinity. This is why the group was not found in the near IR. Images of this region were not included in the GLIMPSE survey and the WISE images do not have sufficient resolution for identification of a stellar group in the vicinity of this YSO. In the case of IRAS 18174-1612, the situation is somewhat different. We were unable to identify it with a stellar object. Its source is presumably a very young and deeply embedded protostar which could only be detected at longer wavelengths or the source could be of extragalactic origin.

We note that there is no significant difference between the distances and the values of the interstellar absorption which, to a certain extent, is also characterized by the value of $\alpha_{IRAS}$, for the range in which the clusters have been detected.

The radii of the detected groups and the surface stellar densities have significant scatter and lie in the ranges of 0.3-2.7 pc and 4-1360 stars/pc$^2$, respectively. The latter value of the density in the IRAS 20198+3716 corresponds to a near estimate for the distance (0.9 pc). It is much higher than the density in the other groups. We may assume that the second estimate for the distance (5.5 pc) corresponds better to reality. The distribution of the surface density for the detected groups corresponds to the peak in the distribution of the density of low-mass

PMS objects at relatively close distances (<500 pc) for young clusters [28].

It is known that the coordinates of the IRAS sources have substantial errors, which makes correlation with other wavelengths difficult. Thus, for identifying the YSOs associated with them, we have used corrected coordinates of the IRAS sources from Ref. 21. Table 2 shows clearly that in the overwhelming majority of cases the distance from the IRAS to the most probable (with respect to its photometric parameters) YSO is considerably smaller; this increases the accuracy of identification.

The identified YSOs are far from always located at the center of the discovered cluster (see Fig. 3). This can presumably be explained by the gradient in the density of the interstellar medium in the cluster, which significantly distorts the real picture.

In 11 of the 18 identified objects (~60%) the IRAS sources are associated with a pair or even a group of YSOs.

If there is no significant difference in the distance and interstellar absorption among the groups discovered in the different IR bands, then some difference in the evolutionary status of the central object may be overlooked. Only among the central objects in groups identified in the near IR are there any PMS stars which can be regarded as objects of a later evolutionary class II relative to their *JHK* photometric parameters. Four of these are in substructures that form part of a more extended stellar cluster.

Given the above results, we may conclude that at least for the example of these 19 regions, groups of young stars should form around YSO in a certain stage of evolution and, with modified selection criteria (depth of images, longer wavelenth range), the percentage of detected groups should be greater. A similar result was obtained for a group of 20 nearer-lying clusters [29].

The search for and discovery of compact star formation regions can be regarded as just an initial stage of research. There is a whole series of questions that will require a more detailed approach and more extensive data bases: In what evolutionary stage of a YSO do compact groups form in its vicinity? Does the richness of a cluster depend on the mass and age of the central object as, for example, in clusters concentrated around Ae/Be Herbig stars [30]? How do the properties of clusters depend on the parameters of the surrounding medium?

This work was supported by a grant to the Byurakan Astrophysical Observatory from the National Academy of Sciences of the Republic of Armenia, "Search for and study of compact stellar groups."

We thank the authors of the UKIDSS GPS [10] and Spitzer GLIMPSE [11] surveys and the authors of Ref. 8.